\documentclass[traditabstract]{aa}

\usepackage{aas_macros}
\usepackage{amssymb}
\usepackage[svgnames]{xcolor}
\usepackage{threeparttable}
\usepackage{graphicx,rotating,epsfig,color}
\usepackage{times,latexsym,txfonts,natbib}
\usepackage{longtable,lscape,multirow,supertabular,subfigure}

\newcommand  \HII{\,H\,{\footnotesize II} }
\newcommand  \HIInospace{\,H\,{\footnotesize II}}

\newcommand  \m{\mathrm}

\bibpunct{(}{)}{;}{a}{}{,} 
\begin{document}
\title{Radiation-pressure-driven dust waves inside bursting \\ interstellar bubbles}
\subtitle{}
\author{B.B. Ochsendorf\inst{\ref{inst1}}, S. Verdolini\inst{\ref{inst1}}, N.L.J. Cox\inst{\ref{inst2}}, O. Bern\'{e}\inst{\ref{inst3}}, L. Kaper\inst{\ref{inst4}} \& A.G.G.M. Tielens\inst{\ref{inst1}}}
\institute{Leiden Observatory, Leiden University, P.O. Box 9513, NL-2300 RA, The Netherlands \\ 
\email{ochsendorf@strw.leidenuniv.nl}\label{inst1}
\and
Instituut voor Sterrenkunde, K.U. Leuven, Celestijnenlaan 200D, bus 2401, 3001 Leuven, Belgium\label{inst2}
\and
Universit\'{e} de Toulouse, UPS-OMP, IRAP, 31028 Toulouse, France\label{inst3}
\and
Sterrenkundig Instituut Anton Pannekoek, University of Amsterdam, Science Park 904, P.O. Box 94249, 1090 GE Amsterdam, The Netherlands\label{inst4}}

\abstract{Massive stars drive the evolution of the interstellar medium through their radiative and mechanical energy input. After their birth, they form ÔbubblesÕ of hot gas surrounded by a dense shell. Traditionally, the formation of bubbles is explained through the input of a powerful stellar wind, even though direct evidence supporting this scenario is lacking. Here we explore the possibility that interstellar bubbles seen by the {\em Spitzer}- and {\em Herschel} space telescopes, blown by stars with log$(L/L_{\odot})$ $\lessapprox$ 5.2, form and expand because of the thermal pressure that accompanies the ionization of the surrounding gas. We show that density gradients in the natal cloud or a puncture in the swept-up shell lead to an ionized gas flow through the bubble into the general interstellar medium, which is traced by a {\em dust wave} near the star, which demonstrates the importance of radiation pressure during this phase. Dust waves provide a natural explanation for the presence of dust inside \HII bubbles, offer a novel method to study dust in \HII regions and provide direct evidence that bubbles are relieving their pressure into the interstellar medium through a champagne flow, acting as a probe of the radiative interaction of a massive star with its surroundings. We explore a parameter space connecting the ambient density, the ionizing source luminosity, and the position of the dust wave, while using the well-studied \HII bubbles RCW 120 and RCW 82 as benchmarks of our model. Finally, we briefly examine the implications of our study for the environments of super star clusters formed in ultraluminous infrared galaxies, merging galaxies, and the early Universe, which occur in very luminous and dense environments and where radiation pressure is expected to dominate the dynamical evolution.}

\authorrunning{B.B. Ochsendorf et al.}
\titlerunning{Radiation-pressure-driven dust waves \\ inside bursting interstellar bubbles}
\maketitle

\section{Introduction}\label{sec:intro}

The morphological appearance of mid-infrared (MIR) bubbles in the disk of the Galaxy is traditionally explained by the interaction of a stellar wind with the surrounding gas \citep{weaver_1977}. In this scenario, stellar winds from massive stars drive shock waves in their surroundings, which sweep up the ambient gas and produce a cavity filled with hot (T $\sim$ 10$^7$ K), tenuous ($n_\m{H}$ $\sim$ 0.01 cm$^{-3}$), collisionally ionized gas surrounded by a dense ($n_\m{H}$ $\sim$ 10$^5$ cm$^{-3}$) shell. Assisted by volunteers from the general public, some 5000 bubbles have now been identified in the Galactic plane \citep{churchwell_2006,simpson_2012}. 

The evolution of expanding \HII\ regions has been addressed numerically in several recent papers, investigating the influence of the initial cloud structure on bubble expansion \citep[i.e.,][]{walch_2013} and the effect of stellar winds \citep{dale_2013}. In addition, a theoretical study including stellar winds is described in \citet{raga_2012}. Still, observations challenge our understanding on wind-blown bubbles (WBB). First, detection of the hot gas through X-ray emission has proven to be elusive for bubbles powered by a handful of OB stars. Diffuse X-ray emission has only been detected towards sources with extreme mass-loss (Wolf-Rayet bubbles; \citet{toala_2012,toala_2013}), extreme ionizing power (M17 and the Rosette nebula; \citet{townsley_2003}), and superbubbles created by interior supernova remnants \citep{chu_mac_low_1990}. Explanations for the lack of X-ray detections inside the vast majority of the bubbles include mass loading, thermal conduction, and leakage of the hot gas from the bubble \citep{harper-clark_2009, arthur_2012}. Second, recent observations show that winds from O stars with log$(L/L_{\odot})$ $\lessapprox$ 5.2 \citep{martins_2005b,marcolino_2009} may be less powerful than long thought (the weak-wind problem; \citet{puls_2008,najarro_2011}), which in its turn is challenged by recent X-ray observations of a stellar wind described in \citet{huenemoerder_2012}. Third, dust grains are expected to evacuate the interiors of WBBs on short timescales ($\sim$10$^5$ yr) by either acceleration or sputtering due to friction with high-velocity gas ($\sim$2000 km s$^{-1}$) from the stellar wind \citep{everett_2010}. Yet, high-resolution infrared observations show that \HII bubbles include a significant amount of dust in their interiors \citep{deharveng_2010,martins_2010,anderson_2012}. The evaporation of small, dense cloudlets could resupply the hot gas with a new generation of dust grains \citep{everett_2010}. While this mechanism can explain the presence of dust inside \HII bubbles, it does not produce in a natural way the morphology of the (mid-)IR radiation: arc-shaped and peaking close to the ionizing source. 

Driven by the difficulties in connection with the model of WBBs outlined above, we have explored the possibilities of \HII bubble formation by thermal pressure \citep{spitzer_1978} instead of a stellar wind. We present the observations and the methods we used in Sec. \ref{sec:method}. In Sec. \ref{sec:bubbleres}, we show that the morphological appearance of dust inside \HII bubbles, blown by stars with log$(L/L_{\odot})$ $\lessapprox$ 5.2), is often characterized by an arc-like structure emitting in the infrared (IR), and propose that these arcs are dust waves \citep{ochsendorf_2014}, where radiation pressure has stalled the dust carried along by a photo-evaporation flow of ionized gas. We test our hypothesis by performing hydrodynamical simulations of the well-known \HII bubble RCW 120 and present a model for studying the coupling between gas and dust, and for predicting the location of a dust wave. We discuss our findings and summarize our conclusions in Sec. \ref{sec:discussion}.
 
\begin{figure*}
\centering
\includegraphics[width=16cm]{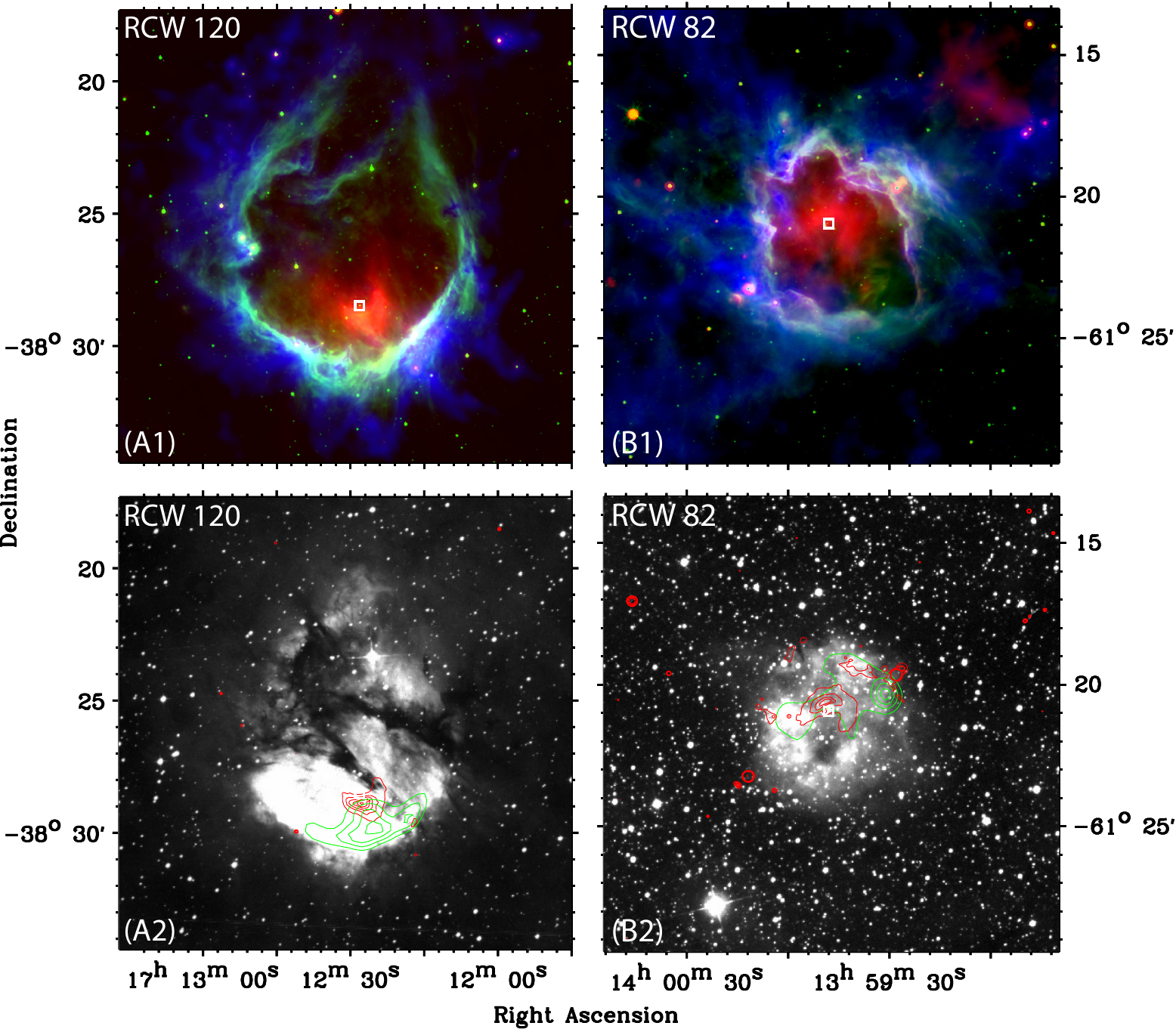} 
\caption{{\bf (A1)} \HII bubble RCW 120 is ionized by a central source (white square) of spectral type O6V - O8V \citep{zavagno_2007,martins_2010}. Spitzer/IRAC observations at 8 $\mu$m (green) trace PAH emission from the inner edge of the bubble, while arc-shaped 24 $\mu$m emission from large $\sim$100 K dust grains, as seen by Spitzer/MIPS, peaks close to the central star (red). The Herschel/SPIRE 250 $\mu$m observations are plotted in blue, revealing high column densities of cold ($\sim$15-30 K) dust surrounding the bubble. Sequential star formation can occur in the swept-up material because the shells are prone to gravitational instabilities \citep{zavagno_2007,pomares_2009,martins_2010}). The shell appears to be broken towards the top side. {\bf (A2)} H$\alpha$ SuperCOSMOS image, overlaid with contours of Spitzer/MIPS 24 $\mu$m (red) and NVSS radio continuum emission at 1.4 GHz (green), revealing the distribution of gas and dust inside the bubble. Contour levels are max - 80\% - 60\% - 40\%. {(\bf B1 \& B2)} RCW 82 is ionized by two stars of spectral type O7V - B2V \citep{pomares_2009}. Images B1 and B2 use the same color scheme as A1 and A2, respectively. The radio continuum data for RCW82 are taken from SUMSS at 843 MHz with contour levels max - 80\% - 60\% - 40\%. In this case, the bubble appears to be broken towards the bottom.}
\label{fig:bubbles}
\end{figure*}

\section{Observations and method}\label{sec:method}

\subsection{Surveys of the Galactic plane}
The Galactic plane has been observed at a wide range of wavelengths, allowing for a multi-wavelength study of the \HII bubbles seen in the GLIMPSE survey \citep{churchwell_2006}. We explored several large-scale surveys that are publicly available online. We summarize these surveys in Tab. \ref{tab:observations}.

\begin{table}
\centering
\caption{Surveys used in this study.}
 \begin{tabular}{l|l|l}\hline \hline
Survey & Passband & Reference \\ \hline
SuperCOSMOS & 0.656 $\mu$m  & \citet{parker_1998} \\
Spitzer/GLIMPSE & 8 $\mu$m & \citet{benjamin_2003} \\
Spitzer/MIPSGAL & 24 $\mu$m & \citet{carey_2009} \\
Herschel/Hi-Gal & 250 $\mu$m & \citet{molinari_2010} \\
NVSS &  1.4 GHz & \citet{condon_1998} \\
SUMSS & 843 MHz & \citet{bock_1999} \\ \hline \hline
\end{tabular}
\label{tab:observations}
\end{table}

\subsection{Hydrodynamical simulations}
Hydrodynamical simulations were produced with FLASH HC (hybrid characteristics), a modified version of FLASH \citep{fryxell_2000} that includes radiative transfer \citep{rijkhorst_2006}. FLASH is a publicly available, modular, parallel, adaptive-mesh-refinement hydrodynamical code. The transfer of ionizing radiation is carried out by the method of hybrid characteristics, which efficiently traces rays across the block-structured adaptive-mesh-refinement grid and computes the intensity of radiation at every computational cell. 
We adopted the on-the-spot approximation, that is, all the ionizing photons produced by recombinations to the ground state were assumed to be absorbed locally, making the radiation transfer equation more simple to solve. Since its first description by \citet{rijkhorst_2006}, the HC scheme has been improved, as described in \citet{raicevic_2010} and \citet{verdolini_2014}. The updated version of the scheme was employed in the radiative-transfer-code comparison project by \citet{iliev_2009}, and tested further for a high-density medium in \citet{verdolini_2014}.

For the simulations presented in this work, we allowed the grid to refine and de-refine according to the second-order density derivative. This choice has the effect of increasing the resolution elements where the density changes on a small spatial scale. Along the edges of the \HII region the resolution increases up to the maximum permitted, which allowed us to properly follow the evolution of the region without an excessive use of computational power.

\section{Results}\label{sec:bubbleres}

Figure \ref{fig:bubbles} presents RCW 120 and RCW 82, which are two well-isolated bubbles, allowing for a detailed study of their morphology and characteristics \citep{zavagno_2007,pomares_2009}. The two bubbles are powered by stars below the so-called weak-wind limit (i.e. log$(L/L_{\odot})$ $\lessapprox$ 5.2): the ionizing source of RCW 120 is a star of spectral type O6V - O8V \citep{zavagno_2007,martins_2010}, while RCW 82 is ionized by two stars of spectral type O7V - B2V \citep{pomares_2009}. The dust properties of both regions have been discussed in \citet{anderson_2012}; their analysis, however, focused on the cold component (10-35 K) seen at long wavelengths (\textgreater\ 70 $\mu$m), which predominantly traces large columns of material of the dense shell. In this work, we focus on the warm dust component, seen at 24 $\mu$m, inside the ionized region of \HII bubbles \citep[see also][]{watson_2008,deharveng_2010}.

The warm dust component at 24 $\mu$m is characterized by arc-shaped emission that peaks close to the ionizing source and reveals that dust resides well within the borders of the swept-up shell, which is traced by 8 $\mu$m emission and appears to be broken or punctured. To study the spatial distribution of gas and compare it with dust, we used high-resolution H$\alpha$ emission data and complemented this with (low-resolution) radio continuum data not affected by extinction in the line of sight. These properties - an arc-shaped structure at 24 $\mu$m and an incomplete shell traced at 8 $\mu$m - are common properties of \HII bubbles \citep{watson_2008,deharveng_2010}, see also Verdolini et al., in prep. We note that ionized gas traces the emission along the line of sight, whereas 24 $\mu$m emission predominantly arises from dust heated to temperatures of $\sim$100 K near the star. This emission could either originate from small stochastically heated particles, often referred to as very small grains (VSGs), or big grains (BGs) in thermal equilibrium with the radiation field \citep[e.g.,][]{paladini_2012}. Whereas several authors attribute the 24 $\mu$m emission inside ionized regions to the increase of VSGs compared with BGs \citep{paradis_2011,flagey_2011}, the analysis of the IR arc in IC 434 revealed that the increased heating by stellar photons from the nearby stars, $\sigma$ Ori AB, can explain the MIR emission \citep{ochsendorf_2014}. Here, we sidestep this problem and note that the emission from 24 $\mu$m and H$\alpha$/radio throughout the bubble interiors shown in Fig. \ref{fig:bubbles} are not spatially correlated: dust emission peaks close to the ionizing source, while the ionized gas does not follow this trend. This implies that gas and dust are (partially; see Sec. \ref{sec:discussion}) decoupled inside RCW 120 and RCW 82. 

\begin{figure*}
\centering
\includegraphics[width=16cm]{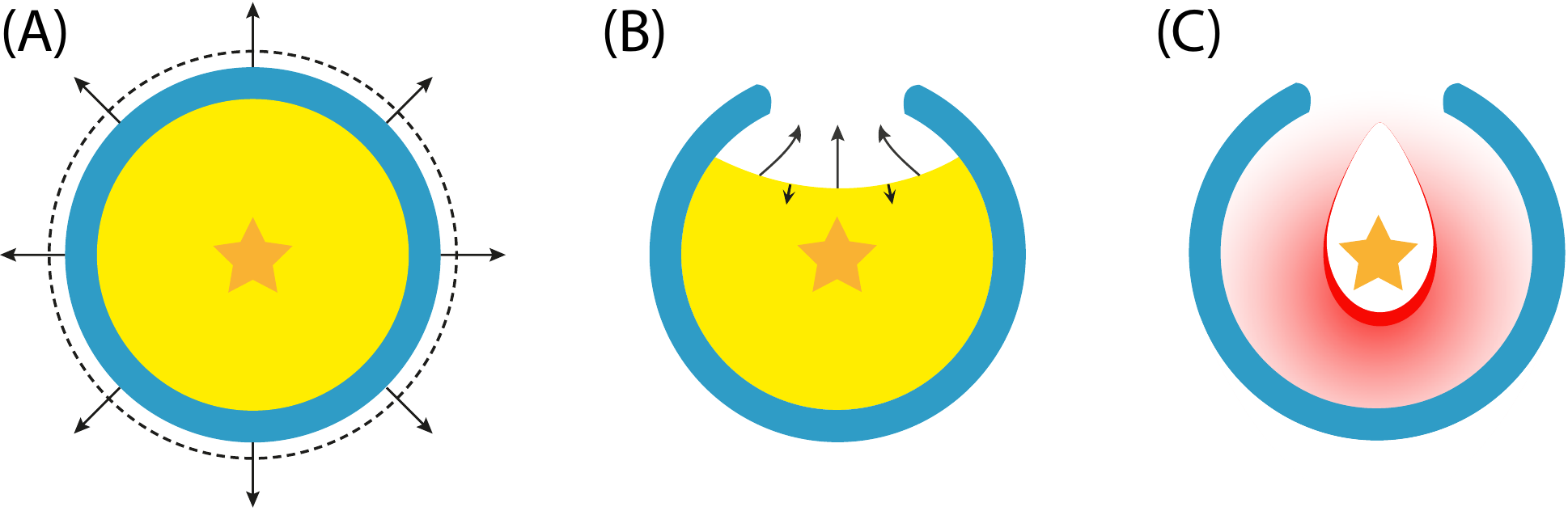} 
\caption{{\bf (A)} Overpressure of the hot, ionized interior (yellow) will cause the bubble to expand inside the natal cloud, sweeping up neutral gas in a dense shell (blue) \citep{spitzer_1978}. If the expansion is supersonic, a shock front forms on the neutral side of the shell (dashed line). {\bf (B)} If the ionized gas contains a density gradient and/or the bubble is punctured, a flow of ionized gas will stream towards lower density and ultimately into the surrounding ISM, relieving the bubble from its pressure \citep{tenorio_tagle_1979}. {\bf (C)} Dust is dragged along in an ionized flow, where 'upstream' dust approaching the ionizing source will be heated and halted by radiation pressure, resulting in a {\em dust wave} or {\em bow wave} \citep{ochsendorf_2014} (see Sec. \ref{sec:location} and \ref{sec:discussion}), which can be traced at mid-infrared wavelengths (red).}
\label{fig:cartoon}
\end{figure*}

We propose that \HII bubbles, blown by stars with log$(L/L_{\odot})$ $\lessapprox$ 5.2), are formed by thermal pressure of the gas that accompanies ionization and not by a stellar wind. The proposed scenario is depicted in Fig. \ref{fig:cartoon}. Using the well-studied bubble RCW 120 as reference, we performed hydrodynamical simulations to test our hypothesis.

\subsection{Hydrodynamical simulations of RCW 120}\label{sec:hydrores}

Stars capable of creating large bubbles are thought to form inside infrared dark clouds (IRDCs), which have typical sizes of 1-3 pc, densities of $\sim$10$^4$ cm$^{-3}$, masses of $\sim$10$^4$ M$_\odot$ \citep{rathborne_2006} and are contained in a larger scale molecular cloud.  We attempted to model the conditions inside an IRDC by constructing a computational domain containing a Bonnor-Ebert (BE) sphere of mass $\sim 7 \times$ 10$^{5}$ M$_\odot$ and radius $\sim$ 9 pc. The observed density profiles of cloud cores are close to those seen in the BE model \citep{Ballesteros-Paredes_2003, pirogov_2009} and, therefore, are often used in cloud-core simulations \citep[e.g.][]{girichidis_2011}. 

A BE sphere is an isothermal gas sphere in hydrostatic equilibrium in a pressurized medium. As initial condition, we set a density profile that follows the BE sphere solution, and did not include unnecessary calculations due to the presence of gravity. We omitted the force of gravity in our simulations because we are only interested in the interaction of the ionizing photons of the star with the surroundings, not in the formation of the star itself. We generated an initial equilibrium solution without gravity by fixing a constant pressure in the computational box and by changing the temperature accordingly. We placed the star $\sim$3 pc {\em offset} from the center of the sphere. Dust is not included in our simulations; however, dust and gas are poorly coupled in evolved \HII regions (see Sec. \ref{sec:discussion}). Dust competes for ionizing photons and will shrink the \HII region in size, which is captured in the uncertainty of ionizing power of the central source and will not affect our conclusions. The number of ionizing photons $Q_\m{0}$ is set to 3.8 $\times$ 10$^{48}$ photons s$^{-1}$ \citep{zavagno_2007}. We followed the evolution of the \HII bubble during the first few Myr of expansion. 

\begin{figure}
\centering
\includegraphics[width=8.75cm]{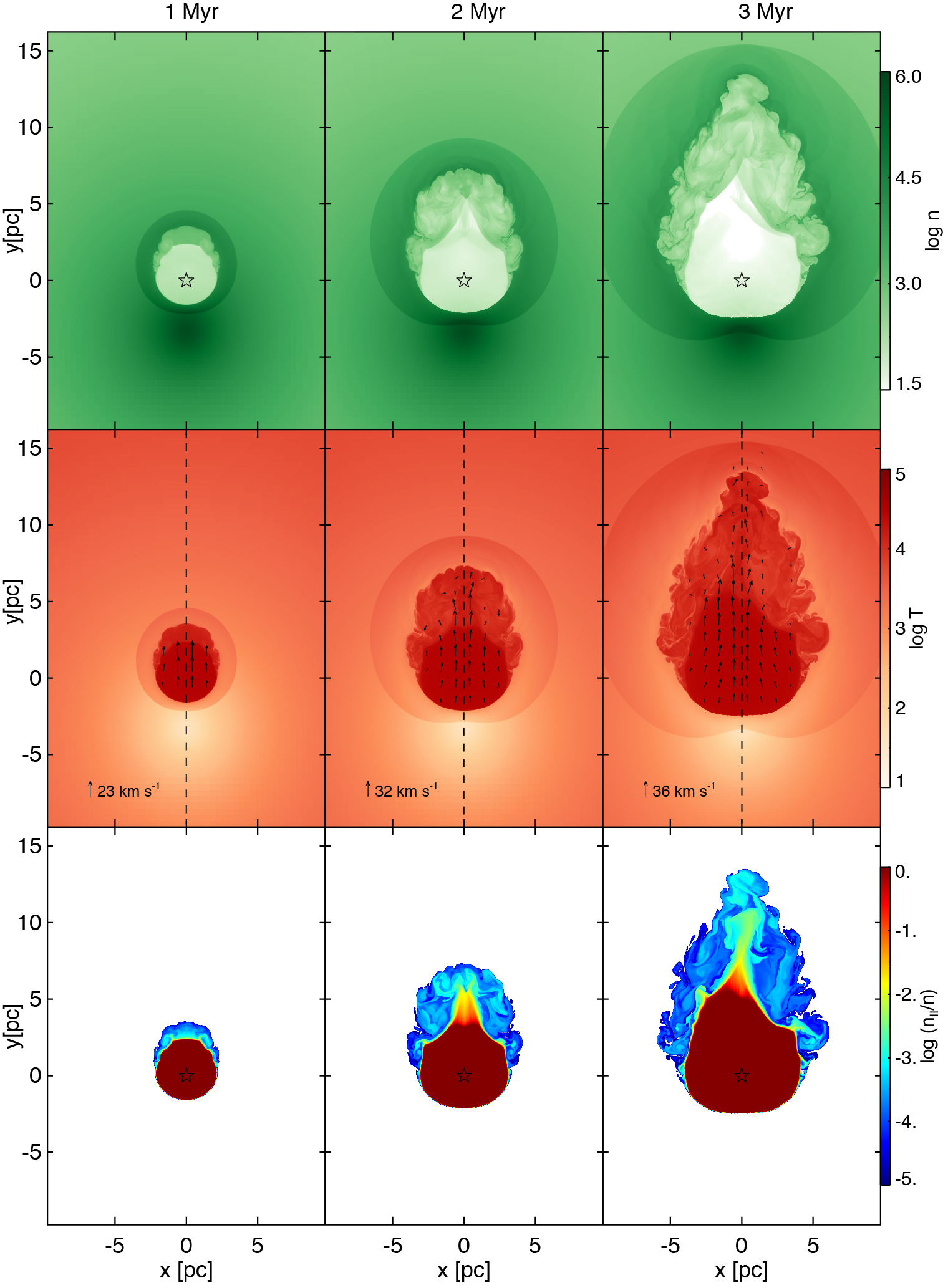} 
\caption{Two-dimensional hydrodynamical simulation of an expanding \HII region offset from the center of a Bonnor-Ebert sphere. From top to bottom row: the logarithm of number density $n$, temperature $T$, and ionization fraction $n_\m{II}$/$n$ of a slice through the $yz$-plane at the location of the star are shown. We trace the velocity of the gas in the three-dimensional simulation along the dashed line (see Fig. \ref{fig:vel_ekin}A). Each column corresponds to a snapshot of the simulation at 1, 2, and 3 Myr. The vectors represent the velocity field in the middle row; the legend indicates the maximum velocity.}
\label{fig:hydrosim}
\end{figure}

\begin{figure}[]
\centering
\includegraphics[width=8.75cm]{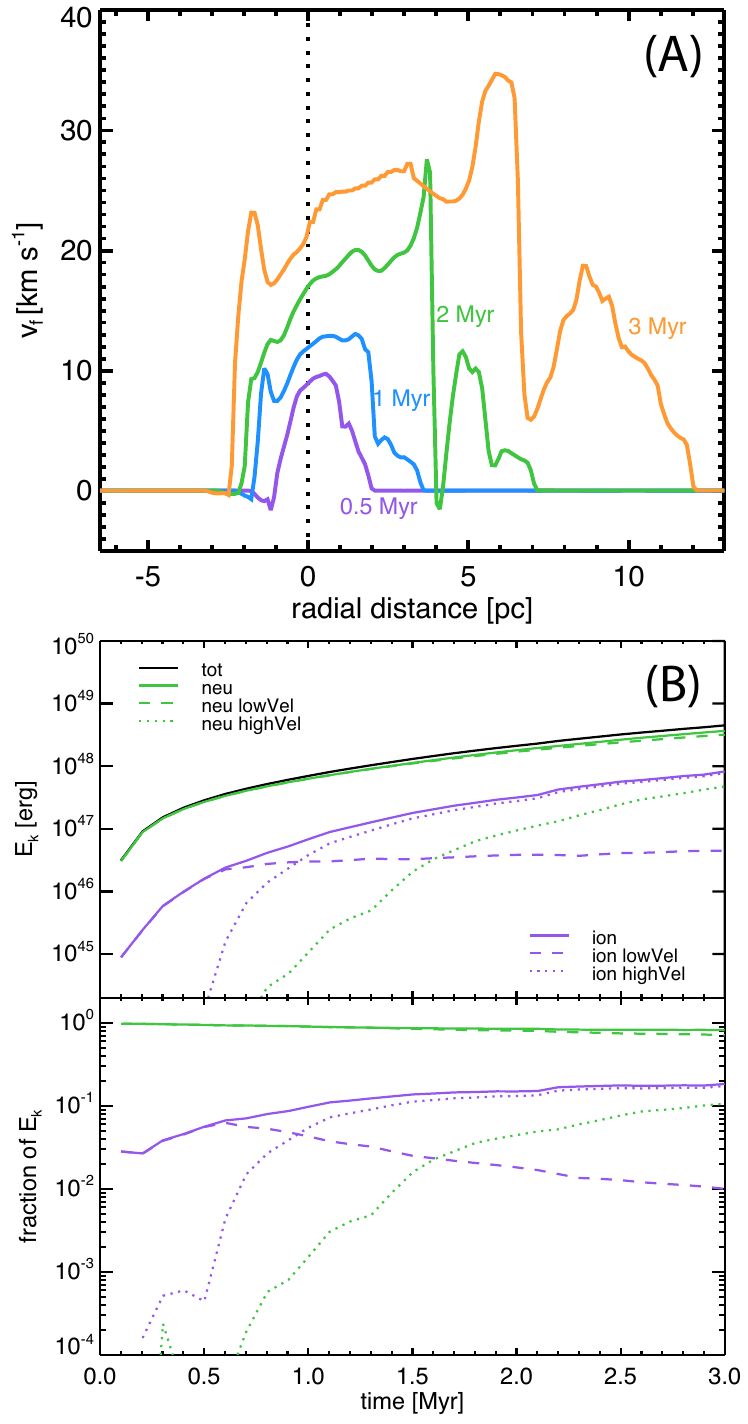} 
\caption{{\bf (A)} Gas velocity along a line in the three-dimensional simulation, which runs from the star through the opening of the bubble (as exemplified in Fig. \ref{fig:hydrosim}). Plotted are the velocity profiles at different times $t$. The star is located at the dotted line. The sharp increase of the velocity close to the ionization front at $t$ = 1 Myr and $t$ = 3 Myr are turbulent motions induced by the finite grid of the simulation. Velocities of the gas with respect to the star range from $v_\m{f}$ = 5 to $v_\m{f}$ = 20 km s$^{-1}$. {\bf (B)} Kinetic energy $E_\m{k}$ contained in the three-dimensional simulation (top panel) and fractional distribution of $E_\m{k}$ (bottom panel). We separate $E_\m{k}$ located into the neutral gas (green), the ionized gas (purple) and their total (black line), which are further divided into the subsonic (dashed) and the supersonic (dotted) part of each component.}
\label{fig:vel_ekin}
\end{figure}

Fig. \ref{fig:hydrosim} shows the results of a two-dimensional simulation with a maximum refinement level equal to 6. Initially, the star creates an expanding sphere of ionized gas, surrounded by a thick shell of neutral swept-up material \citep{spitzer_1978}. However, the density gradient of the BE sphere causes the expansion to accelerate towards low density immediately after ionization ($t$ \textless\ 1 Myr), initiating a photo-evaporation flow inside the ionized gas. Eventually (1 \textless\ $t$ \textless\ 2 Myr), the swept-up shell reaches the edge of the sphere; gas inside the bubble is accelerated to 15-20 km s$^{-1}$. After $t$ \textgreater\ 3 Myr, the shell breaks open towards the low-density region, while the ionized gas inside the \HII region reaches velocities of about 40 km s$^{-1}$. As the so-called champagne flow phase develops \citep{tenorio_tagle_1979}, the relative velocity between the ionized gas in the flow and the star increases (Fig. \ref{fig:vel_ekin}), creating the conditions required for the development of a dust wave (see Sec. \ref{sec:location} and \ref{sec:coupling}). We note that the velocity of the flow, $v_\m{f}$, depends on the starting conditions of the simulations; in particular, the offset of the star from the cloud center and the steepness of the density gradient in the parental cloud, where higher velocities are obtained within a lower density or a steeper gradient. This will be thoroughly addressed in a forthcoming paper (Verdolini et al., in prep).

Fig. \ref{fig:vel_ekin} shows the velocity of the ionized gas and the amount of kinetic energy ($E_\m{k}$) transferred to the surrounding ISM, calculated through a three-dimensional simulation with a maximum refinement level equal to 5. We saved computing power by running the simulation at a lower refinement level than in the two-dimensional case shown in Fig \ref{fig:hydrosim}. This change in resolution does not affect the result of the run, because the energy budget of the simulation does not depend on the refinement level \citep{verdolini_2014,freyer_2003}. We traced the amount of energy in sub- (\textless10 km s$^{-1}$) and supersonic (\textgreater10 km s$^{-1}$) components of the ionized and neutral gas, respectively. Our calculations show that initially, most of the energy is in the neutral shell of swept-up gas. As the neutral shell is accelerated, $E_\m{k}$ is transferred into the neutral gas moving supersonically. The total energy deposited rises steadily as the expanding shell sweeps up ambient material, reaching $\sim$10$^{50}$ erg over the lifetime of the star ($\sim$10 Myr).

\subsection{Model for dusty photo-evaporation flows}\label{sec:location}

The density gradient inside the bubble shown in Fig. \ref{fig:hydrosim} leads to a flow of ionized gas towards lower density. The hydrodynamical simulation did not contain dust; however, in reality, dust will be contained inside the flow and be coupled to the gas through gas-grain interactions. The motion of a dust grain contained in a photo-evaporation flow can be calculated by solving a set of coupled differential equations, including the equation of motion, which balances the radiation pressure force from the star with the drag force through collisions with the gas. Here, we write the relevant equations used in this work, but for a detailed description of the physics of a dusty photo-evaporation flow we refer to \citet{ochsendorf_2014}.

The equation of motion for dust is written as:

\begin{equation}
\label{eq:eqmotiondust}
m_\m{d}v_\m{d}\frac{dv_\m{d}}{dr} = -\frac{\sigma_\m{d} \bar{Q}_\m{rp} L_{\star}}{4 \pi cr^2} + F_\m{drag},
\end{equation}

\noindent where $v_\m{d}$ is the velocity of the dust, and the first term on the right-hand side represents the radiation pressure $F_\m{rad}$; $m_\m{d}$ is the grain mass; $r$ is the distance to the star; $L_{\star}$ and $c$ are the luminosity and the speed of light; $\sigma_{\m{d}}$ and $\bar{Q}_{\m{rp}}$ are the geometrical cross-section and the flux-weighted mean radiation pressure efficiency of the grain. The collisional drag force $F_\m{drag}$ is estimated through \citep{draine_1979}

\begin{equation}
\label{eq:drag}
F_\m{drag} = 2\sigma_\m{d}kTn_i\frac{8s}{3 \sqrt{\pi} } \left(1 + \frac{9\pi}{64}s_i^2\right)^{1/2},
\end{equation}

\noindent where $k$ is the Boltzmann constant, $T$ and $n_\m{i}$ are the temperature and number density of the gas (of species $i$) and $s_i = (m_i v_\m{drift}^2/2kT)^{1/2}$, with $v_\m{drift}$ the drift velocity, which is the relative velocity of the grains with respect to the gas. The equation of motion of gas consists of a balance between momentum gained (or lost) through a pressure gradient and the momentum transfer through interactions with dust grains,

\begin{equation}
\label{eq:eqmotiongas}
v_\m{g}\frac{dv_\m{g}}{dr} = v_\m{g}\frac{c_\m{s}}{\rho_\m{g}}\frac{d\rho_\m{g}}{dr} + \frac{n_\m{d}}{\rho_\m{g}} F_\m{drag},
\end{equation}

\noindent where $v_\m{g}$ and $\rho_\m{g}$ are the velocity and density of the gas, respectively, and $c_\m{s}$ is the local sound speed. The dust number density $n_\mathrm{d}$ is calculated from the MRN size distribution. The ratio $n_\mathrm{d}$/$\rho_\mathrm{g}$ is not constant, but will depend on $v_\m{drift}$ and follows from the equations of continuity.

Consider a dusty photo-evaporation flow, with initial velocity $v_\m{f}$, approaching a star radially. The radiation pressure of the star $F_\m{rad}$ will act on a dust grain contained within the flow, which in this particular case will cause it to lose momentum (i.e., lower $v_\m{f}$). As the dust is slowed down, it will be pushed through the gas with drift velocity $v_\m{drift}$, transferring momentum towards the gas through the drag force $F_\m{drag}$. Eventually, the dust will be stopped at a point $r_\m{min}$ ahead of the star, where the radiation pressure $F_\m{rad}$ balances the drag force $F_\m{drag}$. If at this point the amount of momentum transfer between gas and dust is insufficient for the gas to be dynamically perturbed, the two components will decouple, resulting in an increase of dust upstream relative to the star as the gas flows along unhindered. Dust grains approaching the star with a non-zero impact parameter will be pushed around the star, resulting in an arc-shaped structure \citep{ochsendorf_2014}. The slow-down of the dust will increase the dust number density through continuity, resulting in a pile-up of dust ahead of the star. This structure has been dubbed a {\em dust wave}, the appearance of which will resemble the situation depicted in Fig. \ref{fig:cartoon}C. Dust waves allow us to test the properties of dust in \HII regions: for example, the study of the dust wave in IC 434 implied that dust and gas in IC 434 are not coupled through Coulomb interactions, which is the reason why the plasma drag and Lorentz force are omitted in Eq. \ref{eq:drag} \citep{ochsendorf_2014}. This unexpected result challenges our understanding of the physics of dust in \HII regions, as the grains are expected to be highly charged and to be tightly coupled to the gas through Coulomb focusing. 

We explored the effects of different relevant astronomical environments on the momentum transfer between gas and dust, which determines whether gas and dust will decouple to form a dust wave, or remain coupled to form a bow wave, where gas and dust flow around the star together \citep{ochsendorf_2014}. To do so, we solved the situation sketched above, where a photo-evaporation flow approaches a star radially from the edge of the \HII\ region, for a range of ionizing luminosities $Q_\m{0}$ and densities $n_\m{H}$.

\begin{enumerate}
\item[-] The ionizing luminosity $Q_\m{0}$ determines the size of the \HII region \citep[][p. 243]{tielens_2005}; the Str\"{o}mgren radius of the \HII region is used as the starting point of the flow. Furthermore, $Q_\m{0}$ sets the magnitude of the radiation pressure force $F_\m{rad}$ through extrapolation of O-star parameters listed in \citet{martins_2005}.
\item[-] The density $n_\m{H}$ regulates the collisional drag force, $F_\m{drag}$, as is shown in Eq. \ref{eq:drag} (we consider collisions with hydrogen only). A higher density leads to more collisions and, consequently, a higher momentum transfer between both components.
\end{enumerate}

\begin{figure}[!h]
\centering
\includegraphics[width=8.75cm]{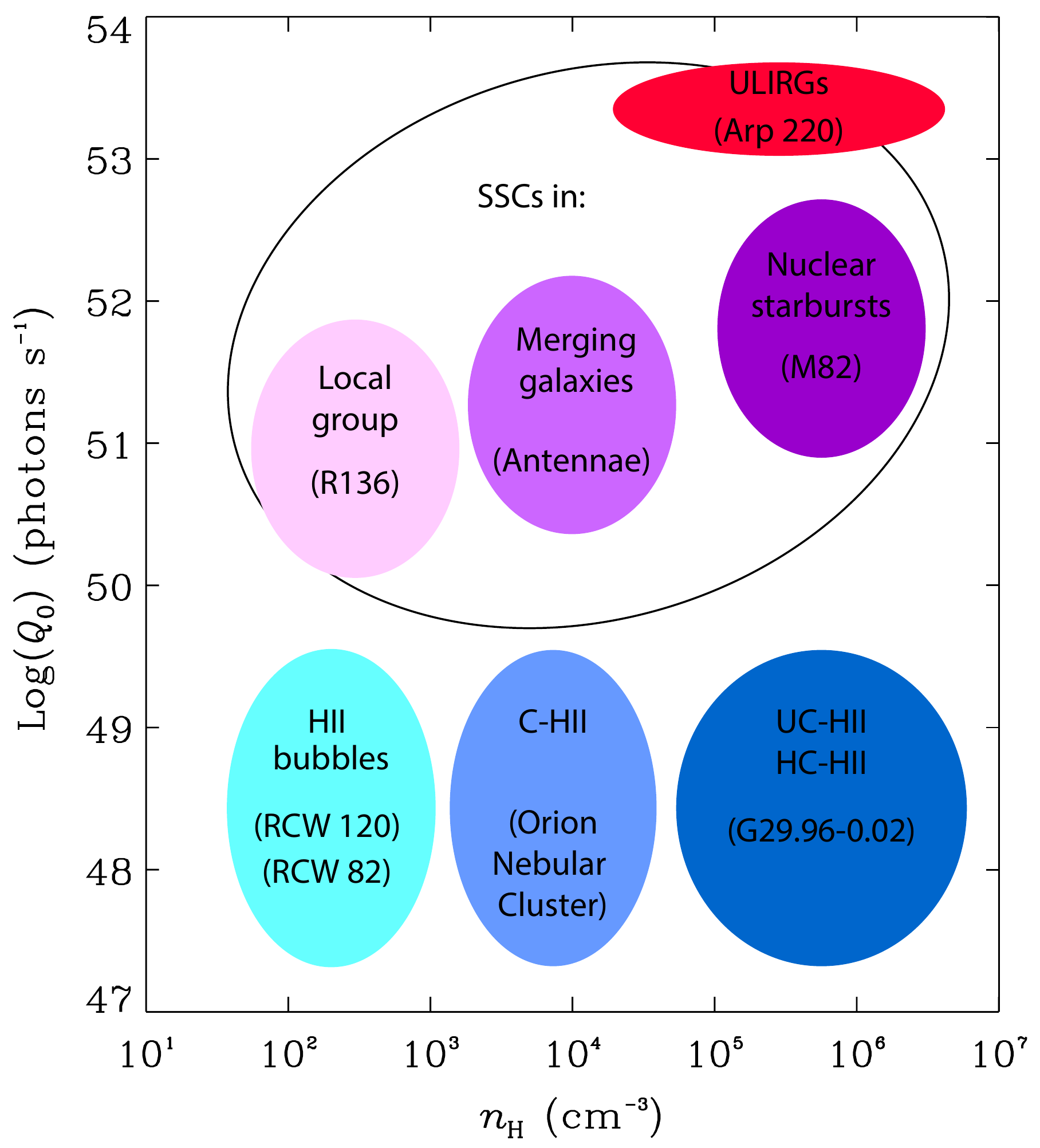} 
\caption{Astronomical environments covered by the parameter space through our choice of the ionizing luminosity $Q_\m{0}$ and density $n_\m{H}$. Among the environments shown are ultra-compact and hyper-compact \HII regions (UC-\HIInospace/HC-\HIInospace), compact \HII regions (C-\HIInospace), and evolved \HII bubbles, as discussed in this work. We also show super star clusters (SSCs) in the Local Group, merging galaxies, starburst galaxies, and ultra luminous infrared galaxies (ULIRGs). Prototypical candidates of each class are denoted between brackets (for references, see Fig. \ref{fig:coupling}).}
\label{fig:environments}
\end{figure}

\noindent We evaluated our model for densities in the range 10 \textless\ $n_\m{H}$ \textless\ 10$^7$ cm$^{-3}$ and for ionizing luminosities in the range 47 \textless\ log($Q_\m{0}$) \textless\ 54 photons s$^{-1}$. Through our choice of $n_\m{H}$ and $Q_\m{0}$, we set up a parameter space that covers a wide variety of astronomical environments, depicted in Fig. \ref{fig:environments}. Furthermore, we calculated the solutions of our model grid assuming several constant flow velocities $v_\m{f}$ of 5, 10, and 20 km s$^{-1}$, which represent typical velocities of ionized gas at the location of the star (Fig. \ref{fig:vel_ekin}A). Note that in reality, gas will accelerate into the \HII region along a pressure gradient (represented by the first term on the right-hand side of Eq. \ref{eq:eqmotiongas}, and seen in Fig. \ref{fig:vel_ekin} through increase of $v_\m{f}$). Much of this acceleration will, however, occur close to the ionization front; adopting a constant velocity is a good first approximation. Our aim is to identify the physical parameters that dominate the momentum transfer between gas and dust, the bulk of which happens close to the star, where $F_\m{rad}$ increases and stops the dust, and we fully expect that the initial acceleration is largely irrelevant for this discussion.

\subsection{Location of dust waves and dust-gas coupling inside a photo-evaporation flow}\label{sec:coupling}

\begin{figure*}
\centering
\includegraphics[width=18cm]{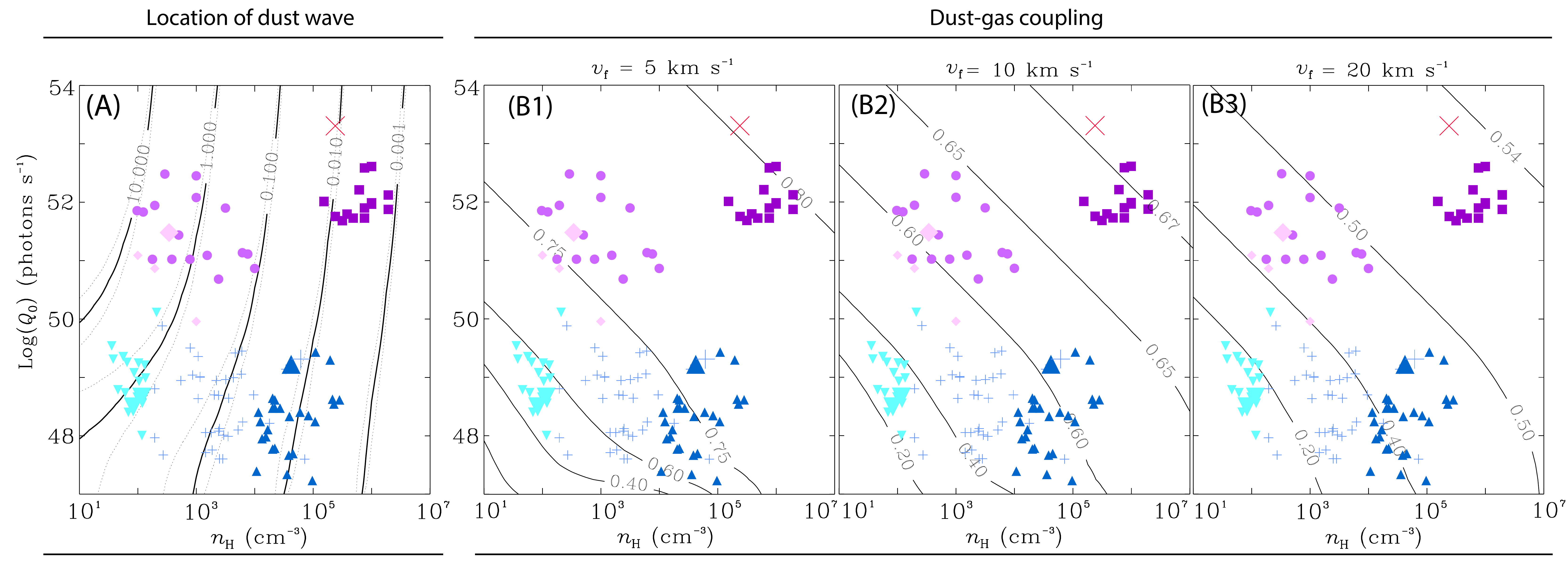} 
\caption{{\bf (A)} Location of a dust wave $r_\m{min}$ (black contours) in pc evaluated for a grid of ionizing source luminosities $Q_\m{0}$ and ambient densities $n_\m{H}$ for $v_\m{f}$ = 10 km s$^{-1}$. The dotted contours represent $r_\m{min}$ calculated with $v_\m{f}$ = 5 km s$^{-1}$ and 20 km s$^{-1}$ (a lower flow velocity will move the contours to the {\em right} in this figure). Overplotted are different astronomical objects, located in regimes drawn in Fig. \ref{fig:environments}: UC-\HII/HC-\HIInospace; $\color{DeepSkyBlue} \blacktriangle$ \citep{wood_1989}, C-\HIInospace; $\color{LightSkyBlue} +$ \citep{garay_1993} and evolved \HII bubbles; $\color{PowderBlue} \blacktriangledown$ \citep{paladini_2012}. We also show SSCs in the Local Group; $\color{LightPink} \Diamondblack$ \citep[compiled from][]{turner_2009}, merging galaxies; $\color{MediumOrchid} \medbullet$ \citep{gilbert_2007}, starburst galaxies; $\color{DarkViolet} \blacksquare$ \citep{mccrady_2007}, and ULIRGs; $\color{Crimson} \times$ \citep{anantharamaiah_2000}. Larger symbols correspond to prototypical candidates of a class denoted between brackets Fig. \ref{fig:environments}. {\bf (B1), (B2), and (B3)} Coupling strength $C$ (black contours), evaluated for a flow velocity $v_\m{f}$ = 5 km s$^{-1}$, 10 km s$^{-1}$, and 20 km s$^{-1}$.}
\label{fig:coupling}
\end{figure*}

Figure \ref{fig:coupling}A plots the location of the dust wave $r_\m{min}$ for 3000 \AA\ silicate grains of specific density $\rho_\m{s}$ = 3.5 g cm$^{-3}$, which is contained in a one-dimensional flow approaching the illuminating star radially from the edge of the \HII region. The size and specific density of the grains set the inertia of the incoming grains with respect to the star and enters Eq. \ref{eq:eqmotiondust} through $m_\m{d} = \frac{4}{3}\pi a^3 \rho_s$, where $a$ is the radius of the grain. We take $\bar{Q}_{\m{rp}}$ = 1 and $n_\m{d}$/$n_\m{H}$ = 1.8 $\times$ 10$^{-13}$ (which is the integrated number density of carbonaceous and silicate dust grains for a bin size of the MRN distribution ranging from 1000 \AA\ to 5000 \AA\ \citep[][p.157]{tielens_2005}). The results in Fig. \ref{fig:coupling}A are shown for several constant flow velocities $v_\m{f}$. 

The dust waves in RCW 120 and RCW 82 are {\em observed} at 0.23 and 0.41 pc \citep{zavagno_2007,pomares_2009}, and are well reproduced with a flow speed of $v_\m{f}$ $\sim$ 10-20 km s$^{-1}$, which is reached in our hydrodynamical simulation after $t$ $\textgreater$ 1 Myr (Sec. \ref{sec:hydrores}). For comparison, RCW 120 is thought to be $\sim$2.5 Myr old \citep{martins_2010}. The exact position of $r_\m{min}$ will depend on the radiation pressure on the one hand and the flow parameters (i.e., the velocity $v_\m{f}$ and density $n_\m{H}$ of the flow) on the other hand.  Fig. \ref{fig:coupling}A shows that the dependence of $r_\m{min}$ on  $v_\m{f}$ is weak. In contrast, $v_\m{f}$ has a strong effect on the momentum transfer from the dust to the gas, as described below. 

The amount of momentum transferred from the dust to the gas depends on the gas density $n_\m{H}$ and on the flow velocity $v_\m{f}$. The ambient number density $n_\m{H}$ sets the number of collisions between gas and dust. At high $n_\m{H}$, the collisions between gas and dust increase in frequency and the amount of momentum transfer to the gas is larger. The velocity of the flow $v_\m{f}$ sets the time $t$ that it takes for the flow to travel the distance from the edge of the Str\"{o}mgren sphere towards the central source. Furthermore, $t$ determines how much momentum can be transferred to the gas over time. Our calculations show that even though fast evaporation flows eventually lead to a higher drift velocity of the grains $v_\m{drift}$ through the gas (i.e., a higher $F_\m{drag}$) and therefore more momentum transfer between the grains to the gas per unit time, the short timescale for a fast photo-evaporation flow to reach $r_\m{min}$ causes the {\em total} momentum transferred to decrease compared with a slow-moving flow of similar density $n_\m{H}$. To summarize, momentum transfer between gas and dust is most efficient in a slow-moving, high-density photo-evaporation flow. 

We have expressed the amount of momentum transfer through the coupling strength parameter $C$, which is a measure of the efficiency of momentum transfer between gas and dust and defined as $C$ = 1 - $p_\m{1}$/$p_\m{0}$, where $p_\m{0}$ and $p_\m{1}$ are the initial momentum of the gas (set by $v_\m{f}$) and the momentum of the gas at the position where dust reaches $r_\m{min}$, respectively. Note that in our model, the dust number density per unit mass of gas (=$n_\m{d}$/$\rho_\m{g}$) is determined through the relative velocity of the gas and dust through the continuity equation and, as such, is not defined at $r_\m{min}$ where the dust velocity $v_\m{d}$ reaches zero. This is a peculiarity of the streamline approaching the star radially: in reality, dust along this streamline will acquire momentum in a random direction because of Brownian motion and flow past the star before the dust is stopped completely. Therefore, we opted to evaluate the momentum transfer up to the point where the dust reaches a velocity to the expected Brownian motion velocity $v_\m{B}$ = $\sqrt{8kT/\pi m_\m{d}}$ = 2.6 $\times$ 10$^{-4}$ km s$^{-1}$). 

If $C$ = 1, dust and gas remain perfectly coupled inside a photo-evaporation flow, and gas will be stopped along with dust. Observationally, a high value of $C$ will express itself in a low contrast between gas and dust and both components would appear to be spatially correlated (a bow wave), similar to the appearance of a stellar-wind bow shock \citep{van_buren_1990,kaper_1997}. When $C$ $\textless$ 1, dust and gas decouple and a higher contrast between dust and gas distribution will be seen in sources with decreasing $C$ (a dust wave). In panels B1, B2, and B3 of Fig. \ref{fig:coupling}, we plot $C$ at different flow velocities $v_\m{f}$. In all cases, $C$ is lowest in evolved bubble \HII regions, which is the main reason that a high contrast between the distribution of gas and dust can be seen inside these sources. However, $v_\m{f}$ has a significant impact on the magnitude of momentum transfer: for example, at $v_\m{f}$ = 5 km s$^{-1}$, $C$ $\sim$ 0.5 for RCW 120 and RCW 82. In contrast, at $v_\m{f}$ = 20 km s$^{-1}$, $C$ $\textless$ 0.2. Again, gas and dust couple relatively well in slower photo-evaporating flows, which could lead to the appearance of a bow wave \citep{ochsendorf_2014}, where dust and gas move around the star together.

\section{Discussion and conclusion}\label{sec:discussion}

The IR emission inside \HII bubbles is often characterized by arc-structures, as exemplified in RCW 120 and RCW 82. We have argued that these structures can be explained as dust waves induced by photo-evaporating flows inside \HII bubbles, which are initiated either by a density gradient inside the bubble or an opening in the bubble shell. The escaping gas will eventually be replaced by a flow of ionized gas evaporating from the inner wall of the swept-up shell. This mechanism provides a natural explanation for the presence and morphology of dust emission seen in the interior of \HII bubbles \citep{deharveng_2010,martins_2010}. Our simulations confirm that an O8 V star can form a bubble of several pc across in typical IRDC conditions without a stellar wind. The bubble is inflated by overpressure of the ionized gas and reveals a similar structure as in RCW 120 and RCW 82, where an ionized gas volume surrounded by a dense shell opens up to one side in order to create a fast champagne flow. 

In this work, we investigated bubble \HII formation in the absence of a stellar wind. As outlined in Sec. \ref{sec:intro}, recent observations have challenged the WBB model \citep{weaver_1977}, given the weak-wind strengths measured for stars with log$(L/L_{\odot})$ $\lessapprox$ 5.2, the difficulty in detecting diffuse X-rays around main-sequence stars, and the presence and morphology of dust inside \HII bubbles. We do not imply that stellar winds are intrinsically absent in these bubbles; the scenario presented here offers an alternative scenario for the formation and evolution of \HII bubbles seen by Spitzer and Herschel because it circumvents the previously mentioned problems. The bubble sample of \citet{beaumont_2010} indicates that the majority of \HII bubbles are indeed blown by stars below the weak-wind limit (i.e. log$(L/L_{\odot})$ $\lessapprox$ 5.2), even after correcting the distance toward the bubbles, which were shown by \citet{anderson_2009} to be systematically higher than the values reported by \citet{beaumont_2010}. However, sources approaching luminosities of log$(L/L_{\odot})$ $\sim$ 5.2 may have a stellar wind contributing to the morphology of gas and dust \citep[see][]{draine_2011}. Moreover, several large bubbles powered by a cluster of stars, with luminosities near to or possibly exceeding the weak-wind limit, also exhibit IR arcs in their interior, for instance, the W5 region \citep{koenig_2008}. In this case, dust and gas seem to follow a similar morphology inside the bubble \citep{deharveng_2012}, which either is the signature of a bow wave or the effect of a stellar wind, as it is unlikely that both the dust and gas will penetrate in the inner region of a WBB dominated by the free-flowing wind \citep{weaver_1977}, and will flow around this region together. More simulations to study gaseous flows inside WBBs will help in understanding the evolution and morphology of bubbles around sources that approach or exceed the weak-wind limit. Nevertheless, dust waves provide direct evidence for the importance of bubble champagne flows, irrespective of the dominant source of expansion (i.e. thermal pressure or stellar winds).

Dust waves require an appreciable flow speed ($v_\m{f}$ \textgreater\ 10 km s$^{-1}$) of the gas near the star for the dust to significantly decouple from the gas. This is reproduced in our models by placing the star offset from the center of a Bonnor-Ebert sphere, whose density gradient leads to a significant acceleration of the gas after ionization of the cloud material. In addition to the acceleration that is needed to segregate the gas and dust to create a dust wave, the density gradient in the cloud also leads to (narrow) openings through which the gas is vented into the ISM, which is characteristic for many bubbles such as RCW 120 and RCW 82 \citep{zavagno_2007,pomares_2009,deharveng_2010}. 

We note that the evolution of interstellar bubbles can be affected by several other mechanisms, such as the motion of the ionizing source, which leads to a cometary shape of the bubble in the direction of movement \citep{wood_1989}, and an external magnetic field, which adds anisotropic pressure to the bubble \citep{bisnovatyi_1995}, possibly leading to elongated shapes aligned along the Galactic field lines \citep{pavel_2012}. This does not seem to apply for RCW 120 and RCW 82, given the near-spherical appearance of the two regions. In addition, we note that grain charging inside \HII regions is not fully understood, and motions of dust grains in IC 434 does not seem to be influenced by Coulomb interaction with the plasma \citep{ochsendorf_2014}. For IC 434, the morphological appearance of the photo-evaporation flow indicates a free flow of the gas and a magnetic field, if present, would then most likely be oriented perpendicular to the cloud surface. Again, it is unclear whether a similar explanation applies to the bubbles studied in this paper. 

Large turbulent instabilities arise once the ionized flow inside the bubble clashes with the dense shell; the signatures of Kevin-Helmholtz and Rayleigh-Taylor instabilities are clearly seen in Fig. \ref{fig:hydrosim}. However, this does not affect the formation of the dust wave contained {\em within} the ionized part of the bubble, because the flow energy of the ionized gas is much higher than the turbulence energy, resulting in a smooth, laminar flow of ionized gas in the interior of the bubble as exemplified in Fig. \ref{fig:hydrosim}. Moreover, gas and dust are only weakly coupled inside evolved \HII regions (see Fig. \ref{fig:coupling}), which minimizes the influence of turbulence on the motion of dust inside a photo-evaporation flow. 

Our models predict that the flow becomes highly supersonic across the bubble, channeling its contents into the surrounding ISM. For the most common type II supernova progenitor (late O/early B-stars with a main-sequence lifetime of $\sim$10 Myr), some 10$^{50}$ erg of kinetic energy can be deposited into the ISM. This is similar to the typical supernova energy ($\sim$10$^{51}$ erg, of which $\sim$10\% is transferred to kinetic energy of the interstellar gas \citep{veilleux_2005}). Moreover, while on a large scale the structure of the ISM is dominated by the walls and chimneys associated with superbubbles created by supernovae breaking into the Galactic halo, Spitzer- and Herschel surveys demonstrate that on smaller scales the structure of the ISM is controlled by the \HII bubbles and photo-evaporation flows discussed here.

The momentum transfer between dust and gas in evolved \HII bubbles such as RCW 120 and RCW 82 is insufficient for the gas to be dynamically perturbed. This makes these objects ideal candidates for observing dust waves and for studying the properties of dust in \HII regions and their dynamical interaction with stellar radiation and gas. However, the velocity of the flow is an important factor: at low velocities, gas remains coupled more efficiently and can form a bow wave. This situation will resemble a stellar wind bow-shock configuration \citep{van_buren_1990,kaper_1997}, where gas and dust peak at a position upstream and produce a drop in emission measure downstream with respect to the star. In regions of high density such as the (ultra)-compact \HII regions, the collision rate between dust and gas increases and the coupling between the two components tightens. Dust and gas also couple well in SSC environments such as R136 in the 30 Dor region and the nuclear starburst in M82. In particular in the latter, dust and gas show similar morphologies near the SSCs that are launching a galactic wind \citep{gandhi_2011}. Radiation pressure is expected to play a key role in the gas and dust dynamics in ULIRGs such as Arp 220. These immense star-forming galaxies will be most efficient in driving gas dynamics through coupling with dust, thereby limiting the efficiency of star formation \citep{andrews_2011}.

Recent studies \citep{krumholz_2009,draine_2011,silich_2013} have explored the importance of radiation pressure for the dynamics of gas around young star clusters. This mechanism is poorly understood, mainly because of our limited understanding of the interplay between radiation pressure, dust, and gas. The study of photo-evaporation flows and dust waves provides us with a unique laboratory to directly study the momentum coupling of stellar radiation and the surrounding medium, which largely proceeds through the dust and is crucial for the implementation in current state-of-the-art models on star and galaxy formation and evolution throughout the history of the Universe.

\begin{acknowledgements} 
 Studies of interstellar dust and chemistry at Leiden Observatory are supported through advanced ERC grant 246976 from the European Research Council, through a grant by the Dutch Science Agency, NWO, as part of the Dutch Astrochemistry Network, and through the Spinoza premie from NWO. NLJC acknowledges support from the Belgian Federal Science Policy Office via the PRODEX Programme of ESA. The software used in this work was in part developed by the DOE NNSA-ASC OASCR Flash Center at the University of Chicago.
\end{acknowledgements}

\bibliographystyle{aa} 
\bibliography{bubbles.bib} 

\end{document}